# The Simultaneous Evolution of Author and Paper Networks

**Katy Börner**

School of Library and Information Science, Indiana University, Bloomington, IN 47405, USA, katy@indiana.edu

**Jeegar T. Maru**

Computer Science, Indiana University, Bloomington, IN 47405, USA, jmaru@indiana.edu

**Robert L. Goldstone**

Psychology, Indiana University, Bloomington, IN 47405, USA, rgoldsto@indiana.edu

**Abstract**

There has been a long history of research into the structure and evolution of mankind's scientific endeavor. However, recent progress in applying the tools of science to understand science itself has been unprecedented because only recently has there been access to high-volume and high-quality data sets of scientific output (e.g., publications, patents, grants), as well as computers and algorithms capable of handling this enormous stream of data. This paper reviews major work on models that aim to capture and recreate the structure and dynamics of scientific evolution. We then introduce a general process model that simultaneously grows co-author and paper-citation networks. The statistical and dynamic properties of the networks generated by this model are validated against a 20-year data set of articles published in the Proceedings of the National Academy of Science. Systematic deviations from a power law distribution of citations to papers are well fit by a model that incorporates a partitioning of authors and papers into topics, a bias for authors to cite recent papers, and a tendency for authors to cite papers cited by papers that they have read. In this TARL model (for Topics, Aging, and Recursive Linking), the number of topics is linearly related to the clustering coefficient of the simulated paper citation network.

## 1. Introduction

Models capturing the structure and evolution of mankind's scientific endeavor are expected to provide insights into the inner workings of science. They are developed to provide objective guidance to augment decisions concerning resource allocation (identification of research frontiers, determining award amount-many small vs. a few large grants), optimum interdisciplinary collaboration (too little collaboration might lead to duplication, too much may lead to rather shallow science), the influence of publishing mechanisms (books vs. fast e-journals), and so on.

Two kinds of models are commonly distinguished: *Descriptive models* that aim to describe the major features of a (typically static) data set and *process models* that model the mechanisms and temporal dynamics by which real world networks (e.g., co-author or paper-citation networks) are created. Most research in *Bibliometrics* (1), *Scientometrics* (2), or *Knowledge Domain Visualizations* (3) has focused on *descriptive models*. For example, research has studied the statistical patterns of co-authorship networks, paper citation networks, individual differences in citation practice, the composition of knowledge domains, and the identification of research fronts as indicated by new but highly cited papers. Recent work in *Statistical Physics* and *Sociology* aims to design *process models*. Of particular interest is the identification of elementary mechanisms that lead to the emergence of small world (4, 5) and scale free network structures (6, 7).

The model proposed in this paper is unique in that it simulates the simultaneous growth of more than one network structure – here authors and papers. The core assumption is that the twin networks of scientific researchers and scholarly articles mutually support one another. Researchers connect articles to one another in co-citation networks, and articles link researchers to one another in co-authorship networks.

The model provides a grounded mechanism for modeling the "rich get richer" phenomenon for paper citation networks as an emergent property of the elementary networking activity of authors reading and citing articles, and also the references listed in read articles. The generalized rich get richer phenomenon is also known as the Mathew effect (8), *cumulative advantage* (9), or *preferential attachment* (10).

The growth of scientific publications and citations is governed by two underlying processes: growth and aging (11). Growth seems to be important for the development of scale free networks. Aging is an antagonistic force to preferential attachment. Even highly connected nodes typically stop receiving links after time has passed. The bias to cite newer papers frequently prevents a scale free distribution of connectivity (12). In the proposed model, an aging bias offsets the "rich get richer" phenomenon for paper citation networks.

A 20-year data set of articles published in the Proceedings of the National Academy of Science (PNAS) is used to validate the model in terms of major network properties of the interlinked co-author and paper citation networks.

The subsequent section reviews related research on descriptive and process models of co-author and paper citation networks. Section 3 discusses desirable features and basic assumptions of the process model introduced in Section 4. Section 5 validates the model by comparing simulated data to a 20-year ISI PNAS data set. Section 6 discusses the influence of model parameters such as the aging of papers in terms of their power to attract citations, the number of topics, and the length of the chain of references that authors consider when making citations. The paper concludes with a discussion and outlook.

## 2 Related Work

There is a long history in bibliometrics (1) and scientometrics (2) of describing the structure and evolution of science (3). As early as 1964, Garfield and his colleagues proposed using citation data to study and write the history of science (13). Citation data has also been used to identify the associations between authors, publications, patents, grants, data and more recently genes, proteins, diseases, etc. Associations have been discovered over time, space, and fields to identify changing frontiers of science (14), to measure science (15), or to recognize research fronts (16).

Research on process models seeks to simulate, statistically describe, or formally reproduce statistical characteristics of interest. Of particular interest are models that "conform to the measured data not only on the level where the discovery was originally made but also at the level where the more elementary mechanisms are observable and verifiable" (17), p.2575.

Recent work in statistical physics aims to design models and tools to analyze the statistical mechanics of topology and dynamics of diverse physical, biological, and social networks. A major goal is to identify elementary mechanisms that lead to the emergence of small world (4, 5) and scale free network structures (6, 7) commonly observed in the real world.

Small world networks have a short average path length among nodes, but a high local clustering coefficient compared to random networks (18). Important small world graph properties include the number of vertices (n), the average degree <k>, the characteristic path length (l) and the clustering coefficient (C). The degree of a vertex is the total number of its connections. The characteristic path length describes how far apart any two nodes in the graph network are. It is computed by determining the shortest path l(i, j) between any two nodes i and j in the network and calculating the average of all l. The clustering coefficient is a more local measure of how "cliquish" a graph is or how tightly nodes in the graph are connected to each other. If a node has K edges that connect it to its neighbors, then the node's clustering coefficient is given by C=N/(K*(K-1)/2), where N is the number of edges connecting neighbors of the node to each other. The strength of a connection (e.g., the number of times two authors wrote papers together) is not considered during the computation of l or C.

In scale free networks, the frequency $f$ of the degree of connectivity $k$ of a vertex is a power function of $k$, $f \sim k^{-\gamma}$. Examples of real-world data sets that are well approximated by power law relationships include actor collaborations, power grids, and the World Wide Web (10). For these data sets, the power law applies over many orders of magnitude, and hence these networks are known as "scale free." With very few highly interlinked nodes and many weakly interlinked nodes, scale free networks are surprisingly robust against random deletion of edges, e.g., network failures (19).

The Watts-Strogatz model was the first model to generate graphs with small world properties (20). Their process model starts with a regular lattice network configuration. Each edge is redirected with a probability p to another randomly selected node. This process model is of limited direct utility for co-author and paper citation networks because in those networks the links are fixed – no rewiring takes place. That is, once a paper has cited another paper, or two authors have collaborated on a paper, these associations are forever part of the permanent historical record.

The Barabási & Albert (BA) model has been a highly influential and successful attempt to simulate networks that show scale free properties (10). It starts with a small number ($N_0$) of nodes, and at every time step, a new node is added as well as a set of m new edges that link the new node to the nodes already present in the system. The probability p that a new node will be connected to node i is proportional to the degree $k_i$ of node i. Hence a new node is preferentially attached to an already highly connected node. After t time steps the network has $N = t+N_0$ nodes and m*t edges. This network evolves into a stationary scale free state with the probability that a node has k edges following a power law with an exponent $\gamma_{BA} = 3$. Gradually adding nodes to the network over time appears to be critical in obtaining scale free distributions (21).

Copying behavior was introduced as an alternative to explain the power law degree distribution for the World Wide Web (22). Recent work by (23) models the probability distribution of citations by copying references used in other papers. The resulting network quantitatively matches the citation distribution observed in real citation networks. Vazquez (24) even suggested that authors do a recursive bibliographic search. In his model a new node is connected to a randomly selected node as well as nodes linked from (referenced by) this node. While these models attempt to capture preferential attachment, less is known about their small world properties. Numerous other attempts to model small world and scale free networks are reviewed in (4, 7).

A number of mathematical models of network evolution have been developed in Sociology. Several models (25) assume a fixed number of edges. Snijders (26) proposed a class of statistical models for longitudinal network data that assumes a directed graph with a fixed set of actors. However, neither the number of nodes nor edges is fixed for evolving co-author or paper-citation networks. The model by Gilbert (27) aims to simulate the structure of academic science. It assumes that papers generate future papers giving authors a rather incidental role. The model was validated based on the number and distribution of citation counts. The small world and scale free properties of the resulting networks are unknown.

To our knowledge there exists no algorithmic approach that simultaneously models the evolution of different networks such as co-author and paper citation networks within an ecology of multiple interacting networks. Here we argue that to fully understand the structure, evolution, and utilization of networks, co-author and paper citation networks need to be considered simultaneously. For example, to understand how knowledge diffuses across authors via their papers at the same time that new authors and papers are accumulated; it is essential to model the coupled growth of both network structures.

## 3  Process Model for Author-Paper Networks

This section motivates the features and simplifying assumptions of a process model for the simultaneous growth of co-author and paper-citation networks as seen in citation databases like PNAS. Given the importance of the interplay of topics, aging, and recursive follow up of links (here citation references) it was named the TARL model.

The TARL model attempts to capture the roles of authors and papers in the production, storage, and dissemination of knowledge. Information diffusion is assumed to occur directly via co-authorships and indirectly via the consumption of other authors' papers. It assumes the existence of a set of authors and papers. Each author and paper is assigned a single topic. Ideally, several levels of topics would be organized hierarchically in terms of specificity.  The same paper may belong to the coarse topic of immunology, the more specific topic of HIV infection, and the still more specific topic of hemolytic anemia in HIV patients with G6-PD deficiency.  The current modeling uses the simplifying assumption that there is a single level of relatively specific topics.  In contrast to the ephemeral lifespan of authors, papers, once written, exist forever.

The set of authors is interlinked via undirected co-authorship relations. Papers are interconnected via directed 'provides input to' links. Authors and papers are interlinked via directed 'consumed' links denoting the flow of information from papers to authors as well as directed 'produced' links representing the act of paper generation by authors. Note that the decision to direct links according to the flow of information reverses the direction of the commonly used 'cited by' link. The in-degree of a paper node refers to its number of references and the out-degree to its number of received citations.

Co-author, citation, consumed, and produced links, once created, are permanent. Co-authorship links may become stronger as more and more papers are co-authored together. The number of 'provides input to' links representing received citations may grow over time. Note that citation links can be created to any existing paper. However, co-authorship links can only be made to currently active authors. For simplicity, each paper has a fixed number of authors and a fixed number of references.

Topics are randomly assigned to authors at initialization time. Papers inherit the topic of their author(s). In the current instantiation of the model, each author and each paper has exactly one topic. The initial number of used topics is typically lower than the total number of available topics. Eventually all the topics are covered as new authors with randomly assigned topics are added gradually. Consumed-produced-relationships among papers and authors are restricted to authors and papers within the same topic. Although this is a rather unrealistic assumption, it parsimoniously models the fact that authors from one knowledge domain typically do not frequently read journals, attend conferences, co-author, or interact from/with other domains.

In the general model, the number of papers produced by an author would be a random variable. However, for the current instantiation of the model, a single fixed number of papers per author per year is assumed. We are aware that this is unrealistic and will not result in a Gaussian or power law distribution for the number of co-authors nor the number of papers published per author. Hence, only the properties of the paper-citation network will be validated against the PNAS data set.

During model initialization, a set of authors and a set of papers with randomly assigned topics are generated (see pseudo-code in Figure 1). Subsequently, a predefined number of co-authors sharing the same topic is randomly selected and assigned to each paper via 'produced by' links. All papers have authors but there may be authors that have not yet produced papers. There are initially no co-author or paper-citation links, making it advantageous to start the model at least one year earlier than the time period of interest.

At each time step (year), a specified number of new authors A' is created with a specific time stamp and added to the set of existing authors $A_t = A' \cup A_{t-1}$. A number of authors can be deactivated as well and subtracted from the set of authors. Subsequently, each author in set $A_t$ randomly identifies a set of co-authors, reads a specified number of randomly selected papers from within his/her topic and produces a specified number of new papers. Each new paper will cite a fixed number of existing papers – a number frequently stipulated by publishers and constrained by the number of pages available per journal. In order to select the papers cited, authors consume (read) a rather small set of papers due to finite cognitive and time constraints. To model the local networking activity of authors, the number of levels of paper references that are followed up by an author is modeled explicitly. If it is zero, then only the papers that an author reads in a year, the set $P_0$, can be cited. If it is two, then an author can cite any paper that they have read, any paper $P_1$ that is cited in one of the $P_0$ papers, or any paper that is cited in any of the papers in set $P_1$. With each deeper level of references, the set of papers, P', is added to $P_r = P' \cup P_{r-1}$. The set of papers available as citation references for a given year t and depth of reference level r is denoted by $P_{r,t}$.

A parameterized model was implemented in Java to simulate the simultaneous growth of interlinked co-author networks and paper-citation networks as described above. The model takes as input parameters that specify the number of authors and papers created in the initial year as well as the number of topics, the number of authors to be deleted per year, the number of papers an author produces per year, the number of papers cited by a new paper, the number of co-authors, the number of levels references are followed up, and the parameters of the aging function.

```
// Initialization
generate #_papers papers and assign a random topic to each paper;
generate #_authors authors and assign a random topic to each author;
randomly assign #_co-authors+1 authors to papers of the same topic;
// Simulation
for each year do {
    add #_new_authors new authors, deactivate authors older than #_author_age;
    for each topic do {
        randomly partition set of authors into author_groups of size #_co-authors+1;
        for each author_group do {
            for each new_paper to be produced, do {
                generate new_paper;
                randomly select #_read_papers from existing papers;
                get all references of read_papers up to #_reference_path_length;
                for each new_paper_reference do {
                    select a time_slice from (start year to curr_year-1) with probability given in aging_function;
                    randomly select a paper published or cited in this time_slice; as a new_paper_reference;
                    add the new_paper_reference to new_paper;
                }
            }
        }
    }
    add all new papers to the set of existing papers;
    add new links to author and paper information;
}
```

**Figure 1.** Process model in pseudo code. If no topics are considered then the number of topics is one, i.e., all papers and authors have the same topic. If no co-authors are considered then each paper has exactly one author. If the reference path length is 0 then no references are considered for citation. If no aging function is given then all papers have the same probability of getting selected.

The model can be started with or without topics, co-authorships, following up of references, aging of papers, or any combination of these variables. A small author-paper network for topics only and for co-authors only is shown in Figures 2a and 2b respectively. Figure 2a shows three unconnected topic-based author and paper networks. During the simulation, authors read, cite, and produce papers from their topic area exclusively. Given that authors exclusively co-author with authors within their own topic, each paper has exactly one topic. Author a4 was assigned a topic for which two papers existed after the initialization and hence both papers are assigned to the author. Later on the author generates one paper each year that cites the author's own work on the topic. Without any new authors or new topic areas for existing authors, all subsequently produced papers will belong to one of these three topic areas. With authors reading papers only of their own topic area, there will never be any links between the three topical clusters. If authors do not co-author then there are no co-author links.

If co-authorship is simulated, then each paper is authored by a predefined number of authors. At each time step, each author will select a number of co-authors to produce papers and each produced paper has multiple authors. The model stops when the number of specified time steps is reached. In the network shown in Figure 2b each newly generated paper has exactly two authors. Blue, undirected lines represent co-authorships. Line thickness indicates the number of papers that have been co-authored together, e.g., a1 and a3 co-authored several times. The total number of papers produced each year is lower than in Figure 2a because two authors produce one paper together. If a topic area has fewer authors than needed for the collaborative production of a paper then no papers are produced.

If no references and no aging are considered then references are randomly selected from the set of papers that a co-author team selected for reading. When references in papers are followed up then authors consider not only the papers they read as potential reference candidates but also papers linked to those via citation references up to a path of a certain length. Thus, a paper that was cited five times has six chances (or tokens) to get selected. The resulting paper-citation network has some nodes, typically older papers, which are very highly cited, while the majority of papers are rarely, if at all, cited; see Figure 3a.

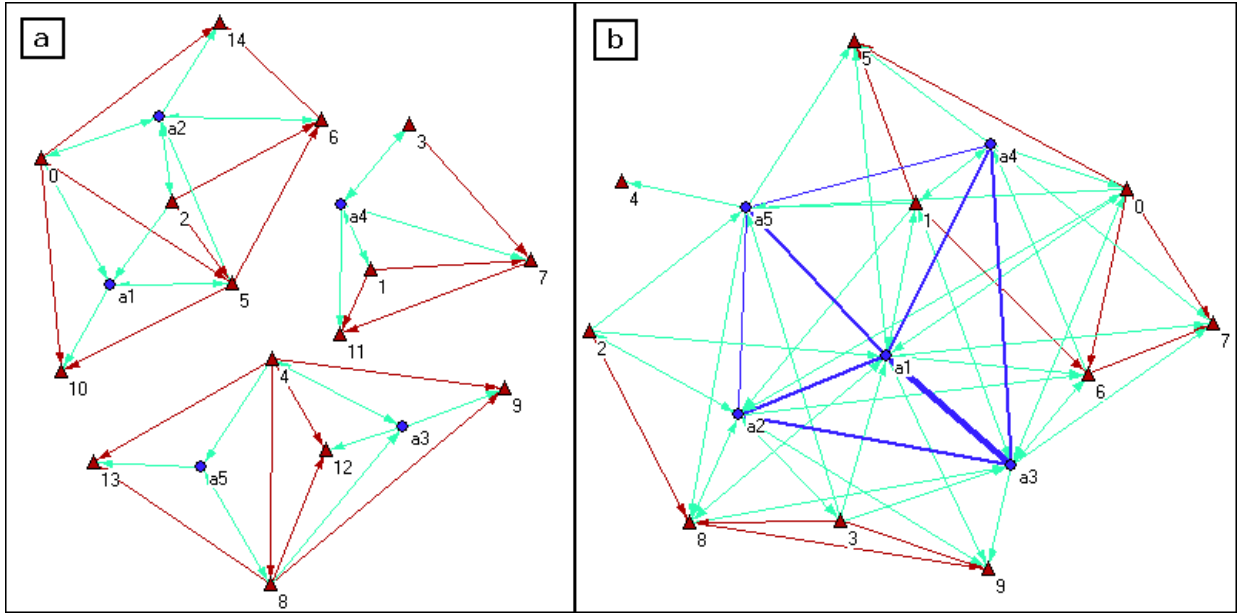

**Figure 2.** Author paper network generated using the model with topics only (a) and with co-authors only (b). The model was started with five topics, authors, and papers, and run for two years. In each year, each author produces one paper, which cites two earlier papers. No authors were added or deactivated. The resulting networks has five authors (labeled a1-a5, blue circles) and 15 or 9 papers (labeled 0, 2, 3…, red triangles). Papers are linked via red directed 'provided input to' links. Authors are connected by blue co-authorships links. Light green directed links denoting the flow of information from papers to authors, and from authors to new papers via 'consumed' and 'produced' relations.

If references as well as aging are considered, then the probability of Paper y being cited, P(y), corresponds to the normalized sum of the aging dependent probability for each of its tokens, $P(y) = \frac{\sum_{t=1}^{n} \sum_{i \in P_{r,t} \wedge i=y} W(t)}{\sum_{t=1}^{n} \sum_{i \in P_{r,t}} W(t)}$, where n is the total number of years considered. Hence a paper that was published in Year y and received 4 citations in Year y+1 and 2 citations in Year y+2 has 7 tokens that are weighted by the probability value for each year. The probability of citing a paper written t years ago can be fit by a Weibull distribution of the form $W(t) = cab^{-a}t^{(a-1)}e^{-\left(\frac{t}{b}\right)^a}$, where *c* is a scaling factor, *a* controls the variability of distribution, and *b* controls the rightward extension of the curve. As *b* increases, the probability of citing older papers increases. For the present purposes, a small value of *b* represents a strong aging bias that favors citing papers that have been published recently. For small values of *b*, the function predicts very few citations for older papers. The introduction of aging offsets the rich-get-richer effect that favors the citation of older papers that have already been frequently cited.

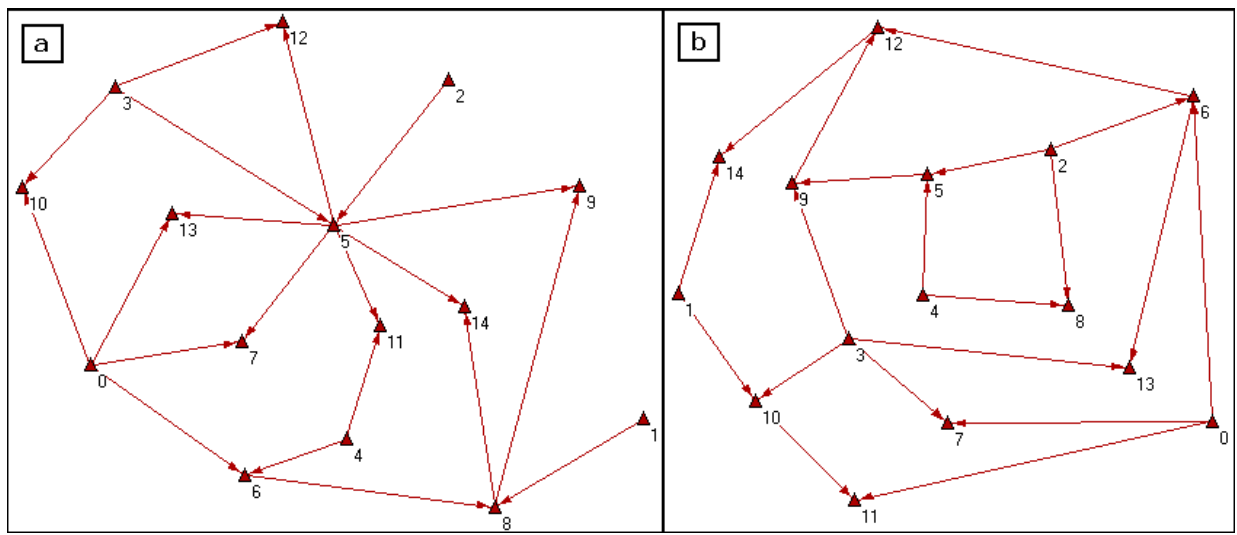

**Figure 3.** Paper network without aging (a) and with aging (b). Without aging, older papers are more likely to provide input to younger papers. That is they are attracting most of the citation links. For example, in a, paper number 0 generated at initialization time and paper number 5 generated in year one provide input to 4 and 6 papers respectively.

The parameters specified in the input script file provide flexibility to fit the model output to diverse data sets. The model is used to fit the PNAS data in Section 4. Section 5 examines the influence of aging, reference path length and number of topics on the structure of interlinked co-author and paper-citation networks.

## 4  Model Validation

To validate the TARL model, a 20-year (1982-2001) data set of the Proceedings of the National Academy of Sciences was used. Subsequently, we describe the data set, select a set of model parameters, and compare the model output with the PNAS data set in terms of network properties.

**The PNAS Data Set**

The PNAS data set contains 45,120 regular articles. The number of unique authors for those papers is 105,915. Table 1 provides counts of the total number of papers, authors, references, and citations received by all of the papers for each of the 20 years, as well as the average number of co-authors per author. The average number of papers published per author as well as the average number of references and citations per paper can be easily derived from the table. Note that the citation counts – particularly for younger papers (green background) – are artificially low because they have not existed in the literature for long enough to garner many citations. The table also provides information on the number of citations received from papers within the PNAS data set in the right-most column. Only those intra-PNAS citation links will be modeled. The paper most highly cited by papers within the set received 285 citations.

**Table 1.**  PNAS Statistics in terms of total number of papers (#p), unique authors (#a), references (#r), citations received per paper (#c), number of co-authors per paper (a#ca) as well as the number of citations (#c_win) within the PNAS data set for each year.

| Year | #p | #a | #r | #c | a#ca | #c_win |
|---|---|---|---|---|---|---|
| 1982 | 1669 | 5201 | 46665 | 56690 | 3.92 | 6749 |
| 1983 | 1611 | 5142 | 46685 | 161437 | 3.98 | 7188 |
| 1984 | 1695 | 5583 | 49834 | 174161 | 4.22 | 6928 |
| 1985 | 1846 | 6325 | 55662 | 191750 | 4.38 | 7425 |
| 1986 | 2042 | 7209 | 64379 | 218229 | 4.76 | 7985 |
| 1987 | 1924 | 7061 | 59110 | 207729 | 4.88 | 7340 |
| 1988 | 2035 | 7471 | 63116 | 215227 | 4.8 | 7547 |
| 1989 | 2088 | 7959 | 65883 | 215437 | 5.01 | 7386 |
| 1990 | 2066 | 8031 | 66019 | 207138 | 5.15 | 7089 |
| 1991 | 2382 | 9559 | 77740 | 223102 | 5.25 | 7511 |
| 1992 | 2500 | 9812 | 80949 | 211238 | 5.29 | 6932 |
| 1993 | 2413 | 9770 | 79848 | 193867 | 5.55 | 5979 |
| 1994 | 2600 | 10656 | 86176 | 187353 | 5.56 | 5910 |
| 1995 | 2476 | 10429 | 82021 | 151249 | 5.66 | 4922 |
| 1996 | 2765 | 11803 | 99061 | 148622 | 5.96 | 5013 |
| 1997 | 2618 | 11255 | 96788 | 122908 | 6.12 | 4290 |
| 1998 | 2711 | 12328 | 100973 | 107764 | 6.48 | 3580 |
| 1999 | 2603 | 12182 | 97018 | 76080 | 6.69 | 2453 |
| 2000 | 2501 | 12201 | 94181 | 44131 | 7.6 | 1354 |
| 2001 | 2575 | 13038 | 97450 | 16357 | 8.4 | 422 |
| **Total** | **45120** | | **1509558** | **3230469** | | **114003** |

Figure 4 visualizes the limited coverage of the data set. It neither contains all work by many authors for the 20-year time span as they may have published in other venues as well, nor does it provide information about citations received from PNAS papers published past 2001 or non-PNAS papers. References to papers outside the 20-year data set will be ignored.

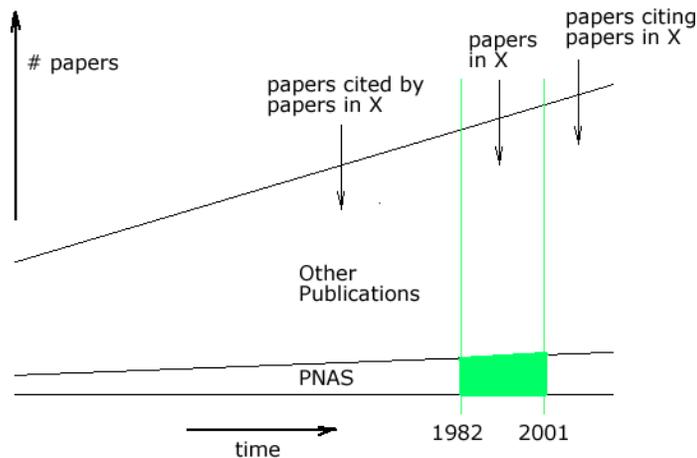

**Figure 4.** Coverage of the PNAS data set in terms of time span, total papers, and complete authors work

Table 2 lists small world properties and power-law exponents for diverse co-author and paper citation networks. The values for the PNAS data set under examination and the simulated paper citation network are also given.

Note that for undirected co-author networks, the in-degree of a node equals its out-degree and hence the exponents for both distributions are identical. For directed paper citation networks, the number of references is rather small and constant. As typical, only the in-degree distribution (received citations) are considered (Dorogovtsev & Mendes, 2002) and the γ reported values for paper citation networks characterize the in-degree distribution. For paper citation networks, we do not report the value for the characteristic path length as it reflects the time duration of the sample but little about the structure of the network.

**Table 2.** Properties of co-author & paper citation networks comprising number of nodes n, average node degree <k>, path length l, cluster coefficient C, and power law exponent γ. Values for the first four co-authorship networks are taken from (28-30). Math and Neuroscience network were analyzed by (21), Redner (31) reported the paper-citation network values for ISI and PhysRev. Values for PNAS and simulated network data were acquired by the authors.

| Network | n | <k> | l | C | γ |
|---|---|---|---|---|---|
| **Co-authorship networks** | | | | | |
| LANL | 52,909 | 9.7 | 5.9 | 0.43 | -- |
| MEDLINE | 1,520,251 | 18.1 | 4.6 | 0.066 | -- |
| SPIRES | 56,627 | 1.73 | 4.0 | 0.726 | 1.2 |
| NCSTRL | 11,994 | 3.59 | 9.7 | 0.496 | -- |
| Math. | 70,975 | 3.9 | 9.5 | 0.59 | 2.5 |
| Neurosci. | 209,293 | 11.5 | 6 | 0.76 | 2.1 |
| PNAS | 105,915 | 8.97 | 5.89 | 0.399 | 2.54 |
| **Paper-citation networks** | | | | | |
| ISI | 783,339 | 8.57 | -- | -- | 3 |
| PhysRev | 24,296 | 14.5 | -- | -- | 3 |
| PNAS | 45,120 | 3.53 | -- | 0.081 | 2.29 |
| SIM | 37,114 | 2.13 | -- | 0.074 | 2.05 |

Based on these values, the PNAS data set can be classified as a medium size data set that has a similar average node degree <k>, path length l, cluster coefficient C, and power law exponent γ to the networks previously examined. The <k> value of the paper-citation network is rather low. The total number of links within the PNAS citation network is 114,003. On average, each paper receives about three citations from another paper in this data set. The co-author network has 472,552 links.

The power-law exponent for the PNAS co-author network is 2.54 and seems to match the values reported for other networks well. It accounts for 91% of the variance in the relation between number of co-authors and frequency of occurrence. The number of authors with very few co-authors is less than predicted by a power-law relation, and the number of authors with a moderate number of co-authors is more than predicted. The best fitting power law exponent for the paper citation network is 2.29, and the power law accounts for 87% of the variance. The systematic deviations from a power law are that most cited papers are cited less often than predicted by a power-law, and the less cited papers are cited more often than predicted. A plausible account for these deviations are that networks in which aging occurs, e.g., actor networks, friendship networks, but also paper citation networks, show a connectivity distribution that has a power law regime followed by an exponential or Gaussian decay or have an exponential or Gaussian connectivity distribution (12). Newman showed that connectivity distributions of co-author networks from astrophysics, condensed matter, high energy, and computer science databases can be fitted by a power-law form with an exponential cutoff (30). Following this lead, we fit a power law with exponential cutoff of the form $f(x) = Ax^{-B} e^{\frac{x}{C}}$. This function provided an excellent fit to the PNAS paper citation network with values of A=13,652, B= .49, and C=4.21 ($R^2$=1.00).

**Model Initialization**

The statistical properties of the PNAS data set were used to select the initialization values for the model introduced in section 3. The model was run with topics, co-authors, references, and aging for 21 years covering 1981-2001. The year 1981 was used for initialization purposes. In 1981, 4809 authors and 1624 papers covering 1000 topics were generated (see discussion of the linear relation between cluster coefficient and topics in section 5). In accordance with the PNAS data, the number of active authors was increased by 430 each year. Note that this increase in authors is mostly caused by external factors such as funding, which are not modeled in the current simulation and hence have to be supplied by hand. Even though 20 years is a rather large time span, the simplifying assumption was made that all authors remained alive/active. Although the number of co-authors increases continuously over time we decided to use the average value of 4. Hence the number of authors per paper is five. One paper is produced by each author per year. The average number of references per paper to papers within PNAS was set to 3 as determined by the actual data. One level of references was considered and the Weibull aging function was used, with a parameter value of $b$=3, providing a 12 year time window in which papers are cited.

**Statistics**

Simulated data (SIM) has been compared to the PNAS article data set in terms of total number of papers (#p), unique authors (#a), and citations received per paper (#c_win) for each year given in Table 1, as well as in terms of their small world properties. Interestingly, the total number of papers in the simulation is slightly lower than the actual PNAS data. This is due to the fact that authors who do not manage to find a sufficient number of co-authors in their topic area will not produce any paper in this particular year. Similarly, papers that are produced in a topic area with very few papers will not be able to reference the called for number of 3 papers. Hence the average degree <k> is slightly lower than the value observed for the PNAS paper network.

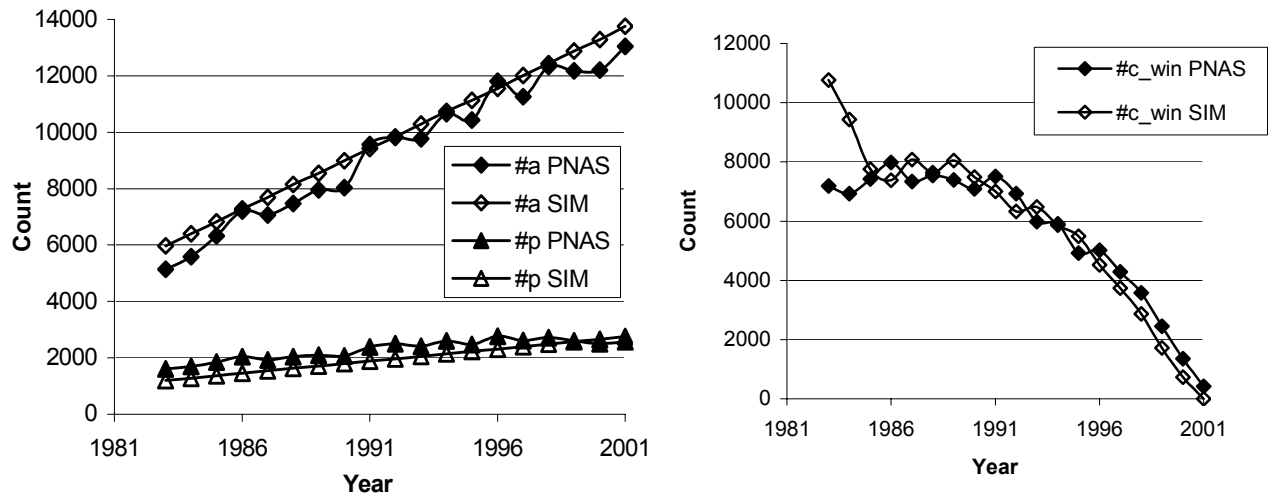

**Figure 5.** Total number of actual and simulated papers (#p), authors (#a) (left) and received citations (#c_win) (right)

As shown in Figure 5, left the number of papers published in PNAS increases slowly but steadily over the 20 year time period. The number of authors publishing in PNAS increases more rapidly than the number of papers because the average number of co-authors per paper increases from 3.49 in 1982 to 5.42 in 2001. The simulation assumes a linear increase in authors over time, but the increase in the number of papers produced naturally comes out of this increase in authors.

The average number of received citations for each year is displayed in Figure 5, right. The model closely tracks the number of actual citations for all years after 1984. The fit for the first two years is poor because the model has no initial citation links nor record of papers prior to 1981. Given that no papers prior to 1981 are available as references, papers published in early years of the simulation receive a disproportionately large number of citations. This effect fades away in 1985 as the aging function selects mostly papers published in the last 12 years and papers published in the last 7 years have a particularly high probability of being cited. The total number of citations received by papers within the PNAS data set is 111,341. The artifacts during the initial phase of the model run could be eliminated by starting the model 10 years earlier and analyzing only the final 20 years. However, we believe the graph on the right of Figure 5 nicely illustrates the influence of aging and the model in action. Both, actual and modeled data sets reflect the fact that younger papers have shorter periods of time in which to draw citations.

**Network Properties**

This section discusses the fit of the simulated paper-citation network to the PNAS data in terms of small world properties as well as the power law exponent γ for the paper connectivity graph. Results for the best fitting model parameters are reported in the last row of Table 2.

The cluster coefficient for simulations in which all authors and papers have the same topic is rather low. The cluster coefficient increases considerably if topics are considered, see also the discussion on topics in the next section.

The simulation with 1000 topics and an aging parameter of $b$=3 provides a good fit to the PNAS data set in terms of the distribution of citations. The model-data $R^2$ was .996, which is substantially better than the best fitting power law to the PNAS data ($R^2$=.87), and almost as good as the best-fitting power law with exponential tail ($R^2$=1.00). As with the PNAS data, the simulated data were fit much better with a power law with exponential tail ($R^2$=.999) than simple power law (.987). Although the simulation does not fit the PNAS data any better than the power law with exponential tail, it does provide a process model for why this functional relation applies. Very highly cited papers are more rare in the PNAS and simulated data sets than predicted by a power law because of the bias toward citing recent papers. The tendency for highly cited older papers to attract still more citations is offset by a counteracting tendency to cite recent papers.

## 5  The Influence of Model Parameters

This section discusses the influence of different TARL model parameters on network properties such as cluster coefficient and power law exponent for the citation distribution.

Interestingly, the number of topics is linearly correlated with the clustering coefficient of the resulting network: C= 0.000073 * #topics. Hence our knowledge about the clustering coefficient in the PNAS network governed the choice of 1000 topics. The linear relation entails a desirable property of the simulation – a simple method exists for creating networks with a specific degree of clustering.

Topics also influence the power law exponent for the citation distribution. Increasing the number of topics increases the power law exponent as authors are now restricted to cite papers in their own topics area. By dividing science into separate fields, the global rich-get-richer effect is broken down into many local rich-get-richer effects, leading to a more egalitarian distribution of received citations.

Aging refers to the distribution of probabilities for papers being cited by new papers. The influence of the $b$ value used to generate different Weibull aging functions is shown in Figure 6, left. The aging distribution observed in the PNAS data was used to determine the parameter value $b$=3, marked with a star in Figure 6. By increasing $b$, and hence increasing the number of older papers cited as references, the clustering coefficient decreases. This effect suggests a second kind of clustering that parallels the strong topic-induced clustering described previously. Papers are not only clustered by topic, but also in time, and as a community becomes increasingly nearsighted in terms of their citation practices, the degree of temporal clustering increases.

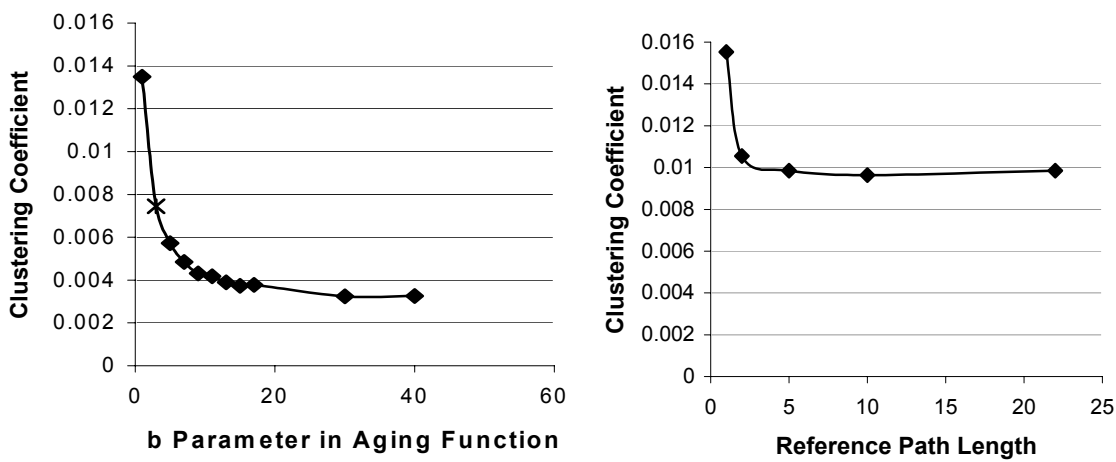

**Figure 6.** Cluster coefficient as a function of the aging function (left) and the reference path length (right)

Last but not least, the length of the chain of paper citation links that is followed to select references for a new paper also influences the clustering coefficient. The dependence of the clustering coefficient on the reference path length is given in Figure 7, right. This result indicates that temporal clustering is ameliorated by the practice of citing (and hopefully reading!) the papers that were the earlier inspirations for read papers.

Note that the aging and reference path length examinations were conducted for 200 topics.

## 6 Discussion and Outlook

This paper presented results on modeling the simultaneous evolution and structure of author-paper networks. Although prior research has described the associations among different scientific structure (e.g., authors, publications, topics, Web) [Menczer, this volume], to our knowledge, nobody has yet attempted to model the simultaneous growth and dynamic interactions of multiple networks dealing with scientific output.

Models based on *preferential attachment* assume that new papers are linked to highly connected (cited) papers and new authors tend to co-author with already highly interlinked authors. However, in today's dynamic scientific world of increasing specialization, an overview of the connectivity of a scientist or paper is not available to authors (even experts in a field). Instead, each author can be seen as a part of a complex network with local connections. Each author interacts directly only with a rather limited number of other authors and papers. However, papers that are cited frequently have a higher probability of being cited again. Similarly, authors that are highly interconnected with other authors in social networks are likely to attract more co-authors if we assume that authors tend to co-author with co-authors of their co-authors. The presented model uses the reading and citing of paper references as a grounded mechanism to generate paper citation networks that are approximately scale free. Moreover, the particular deviations from scale free properties are well predicted by a version of the model that incorporates a bias to cite recent papers and a scientific community that is subdivided into specialized topics. The model parameters that governed these two factors were *b* that reflects that influence of aging and "number of topics" reflecting the degree of splintering within science. The values for these parameters were not freely fit to the citation distribution data. Instead, the number of topics was selected to approximate PNAS's clustering coefficient, and *b* was selected to provide the optimal Weibull fit to PNAS's distribution of citations as a function of lag in years. Thus, the highly respectable model-data fit involving the number of citations and their frequency is impressive because it involves no true free parameters.

The incorporation of topics and recency bias was instrumental in achieving the qualitative violations of a power law distribution. There are fewer papers that receive a large number of citations than is predicted by a power law, because the bias toward citing recent papers offsets the rich-get-richer effect that generates a power law relation. It is difficult for a well cited paper to continue to receive additional citations as it ages.

The citation bias toward recent papers combines with the within-topic citation constraint to create citations networks that have high degrees of clustering by both topic area and time. These model assumptions account for the observed citation distribution, and suggest an interesting interplay between citation practices that lead to egalitarian versus lopsided distributions of citations.

For the sake of simplicity the number of papers produced by each author per year was fixed and a fixed number of co-authors were randomly assigned to each author. If co-authors preferably collaborate with co-authors of their co-authors, this would provide a grounded mechanism for the generation of small world and approximately scale free network structures analogous to their construction in the paper network. Similarly to the aging of papers, the 'deactivation of authors' could also be modeled. If authors are more likely to cite papers of active authors, then the deactivation of all authors of a paper would decrease the 'attraction' or 'fitness' of a paper to receive citations by another paper. The deactivation of authors would also cause previous co-authors to search for new co-authors. Having authors co-author across topics would lead to a more realistic interconnection of papers from different topic areas via citation links.

The productivity of an author may depend not only on his/her position in the author-paper network but also on available research funds, facilities, and students. To give an example, consider the feedback cycle of authors, papers, and funding. Authors that manage to produce many high quality papers also increase their chances of receiving funding. Funding in return enables authors to hire (better) graduate students or post-docs, which in turn increases the number of co-authors as well as the amount and quality of paper output and hence the likelihood of attracting still more funding. Future work will attempt to model grant support as a third network enabling the simulation of this feedback cycle for the first time.

**Acknowledgements**

This work greatly benefited from discussions with and comments from Kevin Boyack, Albert-László Barabási, Mark Newman, Olaf Sporns, Filippo Menczer, and the anonymous reviewers. Mark Newman made code available to determine the small world properties of networks. Nidhi Sobti was involved in the analysis of the influence of model parameter values reported in section 5. Batagelj & Mrvar's Pajek program was used to generate the network layouts (32). This work is supported by a National Science Foundation CAREER Grant under IIS-0238261 to the first author, and a National Science Foundation grant 0125287 to the third author.